\documentclass[reprint, amsmath, amssymb, showpacs, superscriptaddress, aps, prb, nofootinbib
]{revtex4-2}

\usepackage{graphicx}
\usepackage{dcolumn}
\usepackage{bm}
\usepackage[mathlines]{lineno}
\usepackage[usenames,dvipsnames]{color}
\usepackage{subfig}
\usepackage{color}
\usepackage{float}
\usepackage{subfloat}
\usepackage{caption}
\usepackage[]{graphicx}
\usepackage{dcolumn}
\usepackage{bm}
\usepackage{multirow}
\usepackage{url}
\usepackage{gensymb}
\usepackage{setspace}
\usepackage{dsfont}

\usepackage{amsfonts}
\usepackage{booktabs}
\usepackage{siunitx}
\usepackage{dcolumn}
\usepackage[bottom]{footmisc}

\newcommand{\orcidlink}[1]{}

\newcommand{\cmmnt}[1]{\ignorespaces}

\begin{document}

\title{Harnessing the magnetic proximity effect: induced spin polarization in Ni/Si interfaces}
\author{Simone Laterza}
\affiliation{Department of Physics, University of Trieste, Via A. Valerio 2, 34127 Trieste, Italy}
\affiliation{Elettra-Sincrotrone Trieste S.C.p.A. Strada Statale 14 - km 163.5 in AREA Science Park 34149 Basovizza, Trieste, Italy}
\email{simone.laterza@elettra.eu}
\author{Antonio Caretta}
\affiliation{Elettra-Sincrotrone Trieste S.C.p.A. Strada Statale 14 - km 163.5 in AREA Science Park 34149 Basovizza, Trieste, Italy}
\author{Richa Bhardwaj }
\affiliation{Elettra-Sincrotrone Trieste S.C.p.A. Strada Statale 14 - km 163.5 in AREA Science Park 34149 Basovizza, Trieste, Italy}
\author{Paolo Moras}
\affiliation{CNR-ISM, Consiglio Nazionale Delle Ricerche, Area Science Park, Strada Statale 14, km 163.5 Trieste, Basovizza, 34149, Italy}
\author{Nicola Zema}
\author{Roberto Flammini}
\affiliation{CNR-ISM, Consiglio Nazionale Delle Ricerche, via del Fosso del Cavaliere 100, I-00133 Roma, Italy}
\author{Marco Malvestuto}
\affiliation{Elettra-Sincrotrone Trieste S.C.p.A. Strada Statale 14 - km 163.5 in AREA Science Park 34149 Basovizza, Trieste, Italy}
\affiliation{CNR-IOM, Consiglio Nazionale Delle Ricerche, Area Science Park, Strada Statale 14, km 163.5 Trieste, Basovizza, 34149, Italy}
\email{marco.malvestuto@elettra.eu}

\date{\today}

\begin{abstract}

The investigation of the properties of metal-semiconductor interfaces has gained significant attention due to the unique features that emerge from the combination of both metal and semiconductor attributes. In this report, the magnetic properties of Ni/Si interfaces utilizing X-ray magnetic circular dichroism (XMCD) spectroscopy at the Ni and Si edges have been studied. This approach allows to distinguish unambiguously the local magnetism on Ni and Si via individual core-level excitations. Two samples with different semiconductor dopings were investigated using both total electron yield (TEY) and reflectivity configurations.
The experimental results uncovered magnetization at equilibrium in both the metallic layer and in the proximal layer of the semiconductor substrate, implying the presence of induced spin polarization in Si at equilibrium, possibly arising from the depletion layer region. These results hold significant value in the field of spintronics, as similar systems have been demonstrated to generate spin injection through optical medium, opening a new pathway for next generation nonvolatile high speed devices. 

\end{abstract}

\maketitle

\section{introduction}

The integration of magnetic properties into silicon-based materials has been a long-standing goal in the field of Spintronics.
The potential of adding the spin degree of freedom of an electron in charge-based devices is the development of more energy-efficient and faster electronic devices.
However, due to the inherently non-magnetic nature of silicon, the realization of silicon-based spintronic systems has been hampered to date.
Nonetheless, recent experimental and theoretical advancements have opened new avenues to induce and control magnetic properties in silicon, preparing the ground for a new generation of spintronic devices \cite{Zutic04}.
At the forefront of this flourishing field is the design of ferromagnetic/semiconductor interfaces.
These devices already have a crucial role to the current CMOS technology and still continue to offer intriguing insights into both fundamental and applied physics \cite{Tung13, Liu18}. 
\\
Ferromagnetic/semiconductor interfaces are an ideal platform to investigate -and manipulate- the interplay between magnetic and electronic properties of semiconductors \cite{Monch93, Bratkovsky07}.
In fact, these interfaces merge the high magnetic moments, with the tunable electronic properties of semiconductors.
Integrating these characteristics facilitates the conceptualization of innovative devices that exhibit superior performance, heightened energy efficiency capabilities, and advanced designs when juxtaposed with their traditional electronic counterparts \cite{Awschalom07, Wang13}.
Among other things, these interfaces led the discovery of several fundamental spin-dependent phenomena, such as spin injection, spin accumulation, and spin transfer torque \cite{Fert01, Parkin08}.
Spintronic devices as spin valves, magnetic tunnel junctions, and spin transistors \cite{Manchon15} have been based on these phenomena.
Additionally, the investigation of these interfaces has revealed new insights into the role of spin-orbit coupling, exchange interactions, and other mechanisms that govern the behavior of spins at the nanoscale \cite{Awschalom07}.
\\
Recently, a study on the Ni/Si$_3$N$_4$/Si heterostructure has revealed a magnetization dynamics in both the metal and the semiconductor upon optical stimulation \cite{Laterza22}.
The magnetic behavior has been attributed to the generation and propagation of a spin current from the metal to the semiconductor.
Moreover, the equilibrium state of the device clearly exhibits magnetization within the semiconductor substrate, prompting further investigation of the behavior of the proximal silicon layer in the presence of the magnetic field generated by the ferromagnetic film.
\\
In this context, the present research work is focused on this Ni/Si$_3$N$_4$/Si heterostructure, employing X-ray magnetic circular dichroism (XMCD) to probe the magnetic properties of this interface.
The XMCD technique offers the unique advantage of being able to distinguish unambiguously the local magnetism via individual core-level excitations, providing valuable insights into the underlying mechanisms responsible for the observed magnetic behavior.
However, although XMCD is a standard technique to study metallic magnetic systems, limited data is present for semiconductor materials, as for example  the Si L$_{2,3}$, which is investigated only in few Si-based Heusler alloys \cite{Antoniak12, Emmel14}.
The approach in this paper allows for the unequivocal identification of local magnetism on both Ni and the proximal layer of the semiconductor substrate.
By examining two samples with quite different semiconductor dopings by means of total electron yield (TEY) and reflectivity techniques, uncovers the presence of induced spin polarization in silicon at equilibrium, potentially arising from the depletion layer region.
Remarkably, the observed magnetic dichroic signal of silicon is opposite for the two different dopings.
The exceptional interface quality of our samples, as corroborated by the findings from high-resolution transmission electron microscopy (HRTEM) investigations presented in a separate publication \cite{Rajac23}, precludes the possibility that the observed signal emanates from silicide formations.
Further, the contribution from Ni silicides can be ruled out, as they are expected to be nonmagnetic from literature \cite{Dahal16, Ma07}.
This finding is particularly noteworthy as it suggests the existence of a direct magnetic coupling between the ferromagnetic layer and the underlying semiconductor, which has significant implications for the field of spintronics.

\section{Experimental}

\begin{figure}[ht!]
\captionsetup{justification=centerlast}
\centering
\includegraphics[width=0.95\columnwidth]{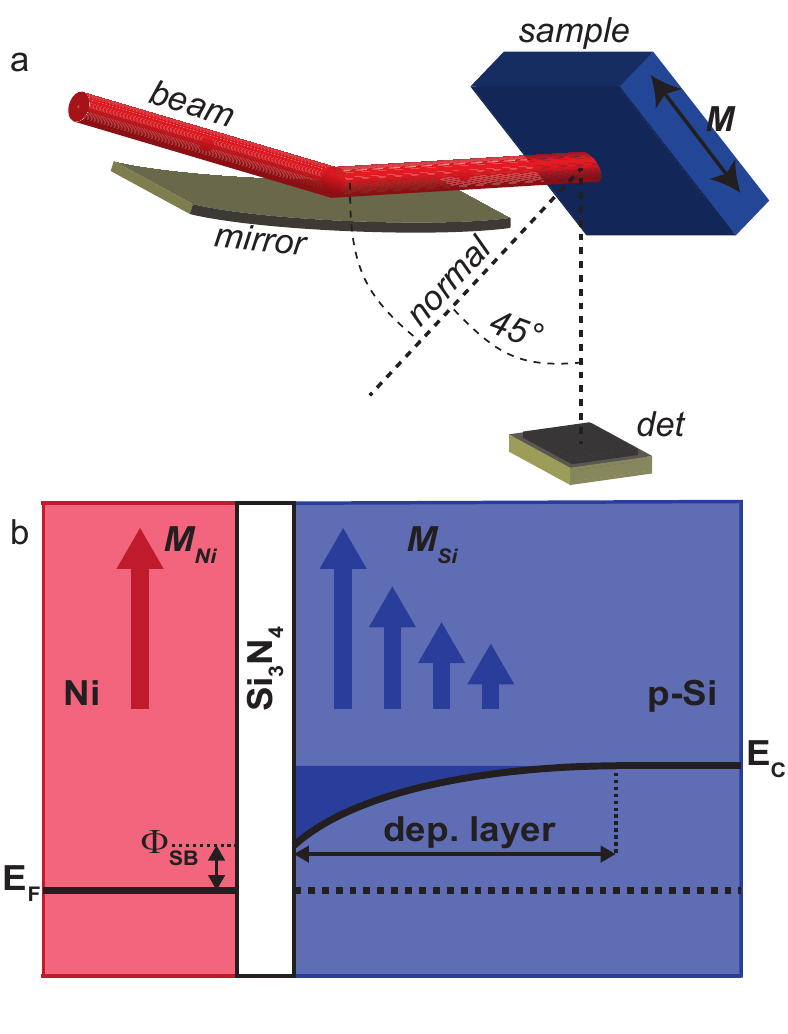}
\caption{\label{fig1}
a) Experimental configuration of the sample with respect to the synchrotron radiation beam and the magnetic field \textbf{B}.
The $I_0$ signal is taken from the toroidal refocusing mirror gold surface.
The angle of incidence is 45$^{\circ}$, while the external magnetic field is along the incoming radiation direction.
A silicon diode detector was positioned at 45$^{\circ}$ from the sample normal in order to collect the reflected radiation.
A schematic representation of the interface is provided in panel (b).
The magnetization of the Ni layer (red arrow) aligns along the surface following the projection of the external field.
The magnetization in the Si substrate (blue arrows) is restricted into  an adjacent layer closer to the surface. 
The lower portion of the diagram displays a sketch of the energy band structure at the Ni/Si interface, highlighting the depletion layer populated by spin-polarized electrons (blue area).
}
\end{figure}

Two Ag/Ni/$\beta$-Si$_3$N$_4$(0001)/Si(111) heterostructures differing only for the doping concentration of the Si substrate - namely low doped (LD), with resistivity 10~$\Omega \cdot$cm\footnotemark[1]\footnotetext{\label{footnote_1} According to the specification of the Si wafer the resistivity of the sample is in the range 1-20 $\Omega \cdot$cm. Here the average value for the resistivity for which the dopant density has been derived is reported.} (N$_d$ $\sim$ $1.5e^{+15} cm^{-3}$ \cite{Masetti83}) and high doped (HD), with resistivity 0.005~$\Omega \cdot$cm  (N$_d$ $\sim$ $2.1e^{+19} cm^{-3}$ \cite{Masetti83}), both p-doped with B dopant - were grown at the VUV-Photoemission beamline (Elettra Sincrotrone Trieste).
The silicon substrates were passivated with nitride in order to reduce the formation of silicides, resulting in the formation of a crystalline bilayer of Si$_3$N$_4$ \cite{Ahn01,Flammini15}, of subnanometric thickness \cite{Rajac23}.
The presence of a finite amount of Ni silicides in the Si substrate is restricted within 3 nm from the Si$_3$N$_4$ layer \cite{Rajac23}.
Subsequently, 7 nm of Ni were deposited at liquid nitrogen to form epitaxial layers \cite{Flammini21}.
Finally, a silver capping layer of 2 nm was grown to avoid the oxidation of the Ni film.
It is possible to estimate the width of the depletion layer for the two samples according to Ref. \cite{Monch93} as $520$ nm for LD\footnotemark[2]\footnotetext{\label{footnote_2} This value has been obtained as the mean integral value for the depletion layer in the range of resistivity reported in note $^{\ref{footnote_1}}$.} and $6$ nm for HD.
A schematic representation of the samples is depicted in Fig.~\ref{fig1}b.
\\
The XMCD measurements were carried out at the CiPo beamline \cite{Derossi95} at Elettra Sincrotrone Trieste, briefly depicted in Fig.~\ref{fig1}a.
The samples were positioned at an incidence angle of 45\degree, enabling the simultaneous acquisition of both the TEY and the reflectivity signals.
The acquired data were subsequently normalized to the intensity $I_0$ of the incident radiation, as determined by the photocurrent measured from the gold coating of the focussing toroidal mirror positioned immediately before to the sample.
The applied magnetic field \textbf{B}, with its direction parallel to the k-vector of the incoming synchrotron radiation, aligns the magnetization \textbf{M} of the Ni film along the line of intersection between the sample surface and the plane of incidence.
The saturating external field during the experiment was set to 200 mT.
Element sensitivity to nickel and silicon is achieved by resonantly tuning the photon energy at the Ni M$_{2,3}$ edge \cite{Koide91} and the Si L$_{2,3}$ edge \cite{Antoniak12}, which are 35 eV apart. 
Throughout the entire range of measurements (50-110 eV), the degree of polarization for the delivered circularly polarized synchrotron radiation is maintained at 90$\%$, the resolving power being 8100. 
All measurements were conducted at room temperature.
\\
The TEY generated from the photoabsorption of the circularly polarized radiation was obtained by means of the drain current from the sample measured with a Keithley 6512.
The reflection from the sample was instead collected by using an AXUV100G silicon diode whose current was measured by means of an Agilent 34401A multimeter.
The signal intensity at each energy point of any spectrum was the result of averaging 5 measurements at an integration time of 300 ms.
Then, a baseline was removed on the pre-edge plateau for each resulting traces by means of a linear fitting, and subsequently the spectra were normalized to a region above the edge.
Finally, the XMCD signals were calculated as the difference of the spectra taken at opposite fields (helicities) while mantaining the helicity (field) fixed.
\\
Additional static magnetic measurements were conducted at the MagneDyn end-station \cite{Svetina16, Malvestuto22}  utilizing the externally seeded EUV free-electron laser (FEL) FERMI \cite{Allaria13} at Elettra Sincrotrone Trieste.
Specifically, resonant magneto-optic Kerr effect (RMOKE) magnetic hysteresis measurements were obtained with the probe resonantly tuned at the Ni M$_{2,3}$ edge and at the Si L$_{2,3}$ edge.
The details on the experimental setup and on the measurements can be found in other sources \cite{Caretta21, Laterza22}.
Additionally, the same interface has been characterized by HRTEM, revealing that the presence of silicides at the interface was kept under control during the sample growth \cite{Rajac23}.

\section{Results}

\begin{figure}[ht!]
\captionsetup{justification=centerlast}
\centering
\includegraphics[width=0.95\columnwidth]{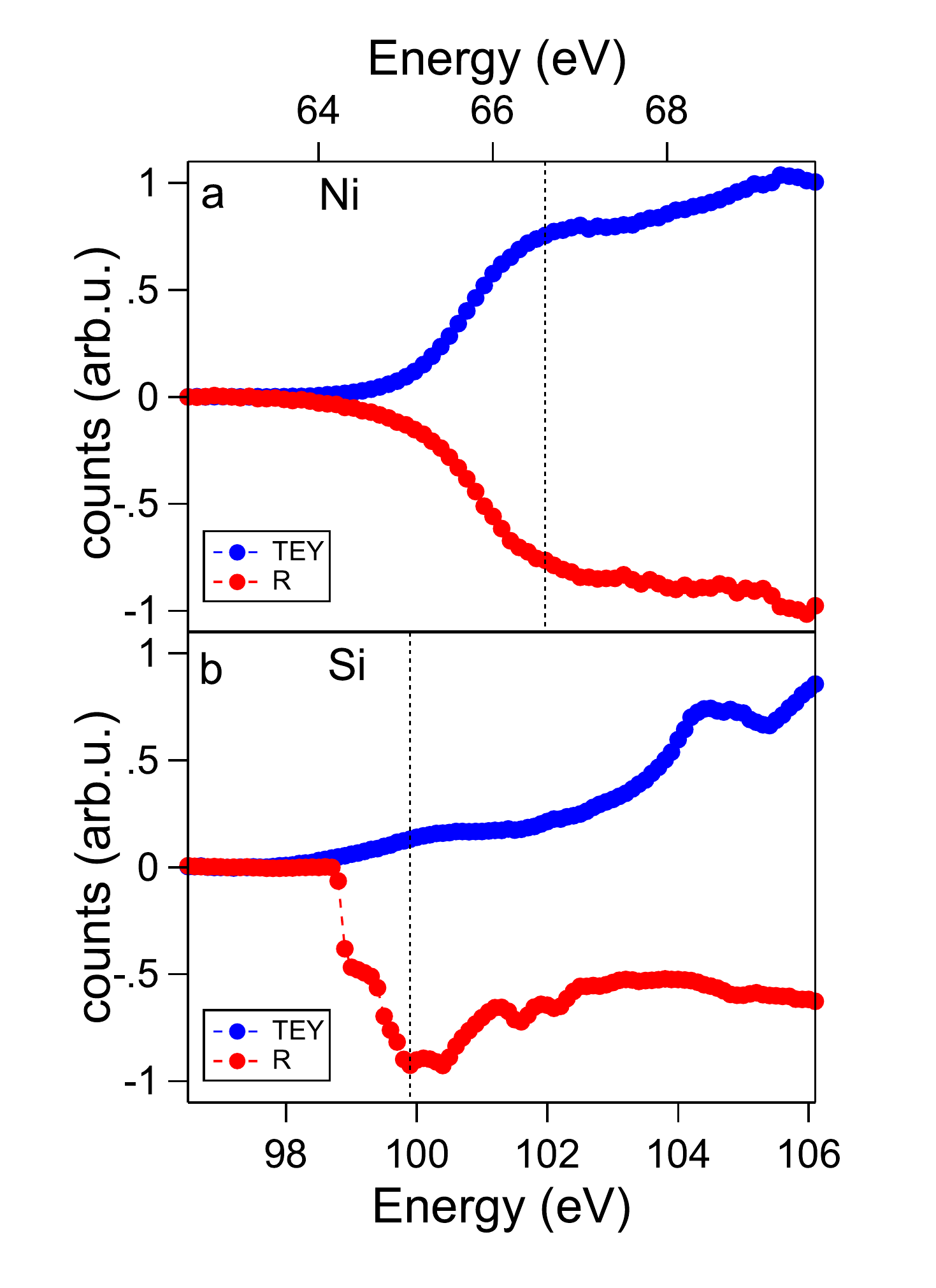}
\caption{\label{fig2}
Reflectivity (red) and TEY (blue) signals at the Ni M$_{2,3}$ edge (panel a) and at the Si L$_{2,3}$ edge (panel b) on LD sample.
The sign of the reflectivity signal has been inverted for clarity.
The black dashed vertical lines in both panels indicate the energies at which the RMOKE hysteresis were collected (see Supplemental I).
}
\end{figure}

Figure \ref{fig2} displays the reflectivity and TEY signals (red and blue curves, respectively) taken a the Ni M$_{2,3}$ edges (panel a), as well as the Si L$_{2,3}$ edge (panel b).
At the Ni edge (Fig. \ref{fig2}a), both reflectivity and TEY lineshapes display a clear increase at approximately 66 eV.
\\
On the other hand, while the reflectivity line shape shows a clear resonance signal at the Si L$_{2,3}$ threshold  (Fig. \ref{fig2}b), this is not the case for TEY.
Consequently, the corresponding XMCD signal is much more evident in the reflectivity signal.
This can be rationalised by considering that the probing depth of reflectivity is much more bulk-sensitive than that of the TEY \cite{Frazer03}.
\\
Hence, even though the magnetic signal is present in both TEY and reflectivity, the subsequent discussion is solely focused on the reflectivity measurements.

\begin{figure}[ht!]
\captionsetup{justification=centerlast}
\centering
\includegraphics[width=0.95\columnwidth]{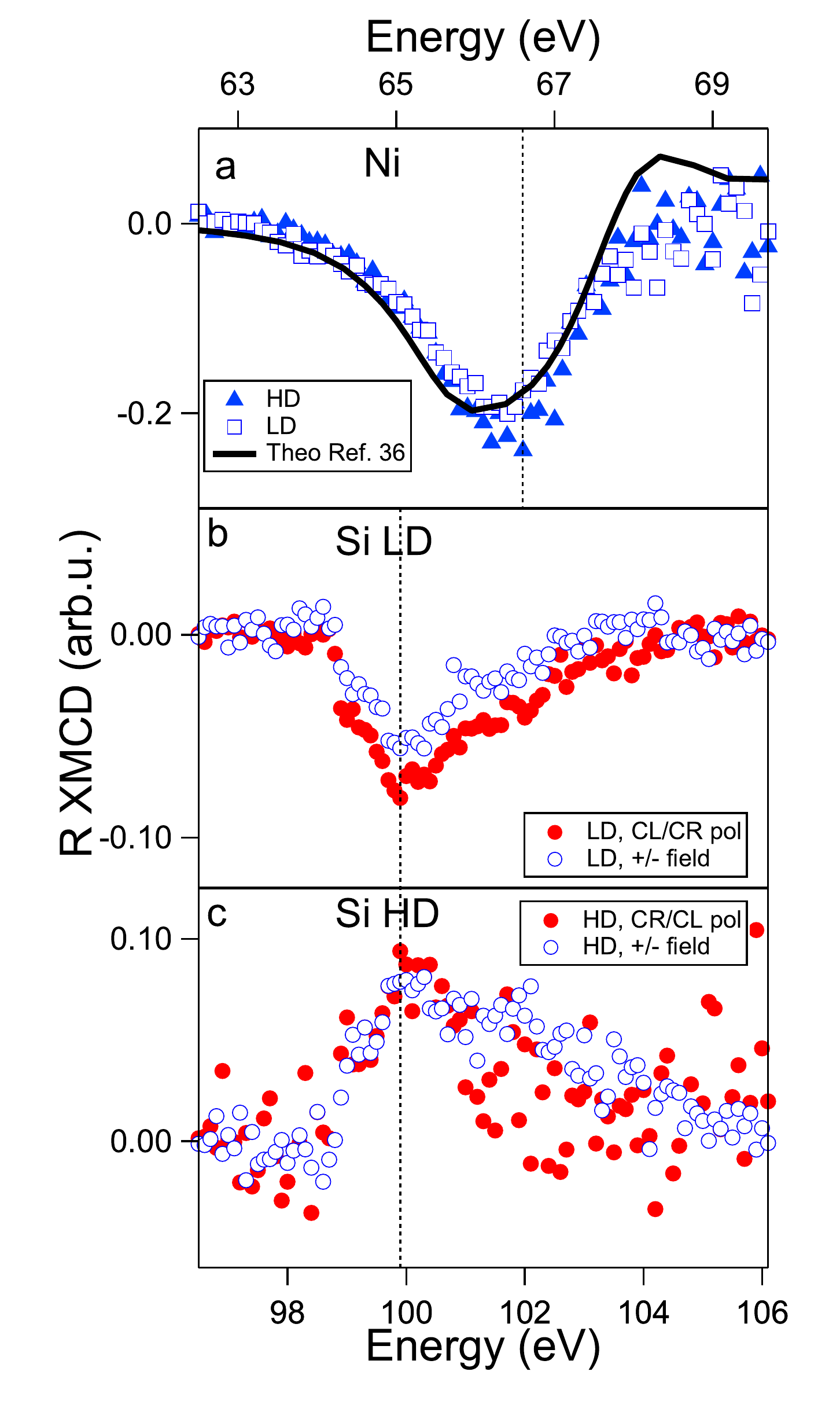}
\caption{\label{fig3}
XMCD signals collected from the Ni/Si heterostructures.
a) XMCD signal in reflectivity at the Ni M$_{2,3}$ edge on LD (squares) and HD (triangles); the complementary behaviour confirm the magnetic nature of the metallic film for both the samples.
Data are compared to the theoretical expectation (black curve) taken from Ref. \cite{Willelms19}. 
b) XMCD signal in reflectivity at the Si L$_{2,3}$ edge on LD collected with the same helicity at opposite fields (blue) and viceversa (red); both the signals reveal a single peak at the Si edge position.
c) XMCD signal on HD; both the XMCD peaks at the Si L$_{2,3}$ edge are reversed in sign.
}
\end{figure}

The Si L$_{2,3}$ absorption edge corresponds to an excitation from the 2p inner shell to a final state associated with the minimum of the conduction band \cite{Brown77} at a threshold of 99.8 eV.
The 2p inner shell is split by a spin-orbit interaction of 0.61 eV, with a statistical intensity ratio of 2:1 between the L$_2$ and L$_3$ components.
In accordance with Refs. \cite{Kasrai96, Hu04, Dorssen98}\cmmnt{for the reflectivity (1-R) - Nayak10 - calculated values}, a clear determination of the origin of the peaks in the line-shape reflectivity spectra at the Si L$_{2,3}$ edge \cmmnt{(Fig.~\ref{fig2}b) }can be made. 
The initial edge signal, which encompasses a pre-edge at 99 eV, a primary peak at 100 eV, and a post-edge feature at 102 eV, can be ascribed to Si$^0$ oxidation state.
In the higher energy domain, specifically within the 104-110 eV range, a secondary peak emerges due to the presence of Si$^{4+}$ nitridation state, which derives from the Si$_3$N$_4$ layer \cite{Lucovsky11, Leitch04}.
It is noteworthy that the established Ni silicides absorption edge position resides within 0.1-0.3 eV above the Si absorption edge \cite{Aballe04, Tam09}.
\\
In contrast, the scenario at the Ni M$_{2,3}$ edge is considerably more straightforward, as the M$_2$ and M$_3$ peaks at positions 65.3 and 66.9 eV \cite{Hochst95, Willelms19} merge into a single peak.
In this context, the transition responsible for generating the XMCD signal takes place from p-states to d-states, while the transition into s-states can be disregarded.
\\
Fig. \ref{fig3} displays the XMCD reflectivity signals collected at the Ni M$_{2,3}$ edge (67~eV) and the Si L$_{2,3}$ edge (100~eV).
In Fig. \ref{fig3}a the XMCD signal at the Ni edge shows a single almost-symmetric peak at 66.5 eV for both the LD (squares) and HD (triangles) samples, as the two metallic films have the same thickness and magnetic state.
In Fig. \ref{fig3}b the XMCD reflectivity collected both at opposite fields at fixed helicity (blue) and at opposite helicities at fixed field (red) at the Si edge for the LD sample is shown.
The signal displays a single negative peak, rising from zero at 99 eV to the maximum amplitude at 100 eV and slowly decaying again to zero above 104 eV.
Figure \ref{fig3}c shows similarly the XMCD reflectivity for HD.
As it is known that the sample aligns the metallic film magnetization along the surface and in the same direction of the external magnetic field, it can be easily deduced that the substrate magnetization of LD is ``ferromagnetic-like``, whereas HD behaves antiferromagnetically, meaning that the Si magnetization is opposed to that of Ni.
Excluding the sign, the shapes of both XMCD reflectivity and RMOKE hysteresis (see Supplemental I) of LD and HD are comparable.

\section{Discussion}

Silicon is inherently a weakly diamagnetic material, and as such, an XMCD signal at any absorption Si edge is not typically expected.
However, the observations reported in Figs.~\ref{fig3} b and~\ref{fig3} c - as well as in Supplemental I - reveal a distinct magnetic signal in the present Ni/Si interfaces.
The two potential effects that could account for the origin of magnetism in silicon, both of which are based on the proximity effect of the silicon substrate in relation to the magnetized Ni layer, have been inspected.
\\
This effect may be attributed to the diffusion of low-energy thermal electrons from Ni to Si.
At equilibrium within the metallic layer, spin-minority and majority electrons experience differing exchange interactions with the predominantly spin-majority polarized electron background.
As a consequence, spin majority scattering times and mean free paths are greater than the corresponding values for minority electrons.
Therefore, thermal electrons impinging on the Ni/Si interface are also spin-polarized and can diffuse in the semiconductor as a tunneling current.
This electron flux must be counterbalanced by an equal flux of electrons from the semiconductor to the metal, since the applied bias is zero, thus preserving the zero net charge current.
The distinct spin polarization of the two opposing fluxes can lead to a net spin accumulation in the proximal layer of silicon \cite{Dankert13}.
\\
A second possible source of magnetism in silicon may be attributed to the proximal magnetic field, which induces a spin polarization for the conduction electrons in the Schottky barrier depletion layer.
Indeed, although the interlayer technically acts as an insulator, the thinness of Si$_3$N$_4$ causes the sample to behave like a sharp metal/silicon interface.
Additionally, the interface forms a Schottky barrier that varies linearly with the insulator thickness, without pinning of the Fermi level  \cite{Sobolewski89}.
The Schottky barrier height is $\Phi_{SB}$ of +0.02~eV \cite{Sobolewski89}, and the depletion layer ranges from approximately 520 nm for the LD sample to approximately 6 nm for the HD sample.
In both cases, due to the presence of the ferromagnetic Ni, electrons in Si will perceive an effective magnetic field that causes exchange splitting and, consequently, a net spin polarization in the depletion layer.
The observation that the two samples exhibit opposite magnetization is particularly intriguing and could suggest that the two different effects are effectively competing with one another.
\\
Further, the role of silicides can be ruled out, as the known phases of nickel silicides are nonmagnetic \cite{Dahal16, Ma07}, and their presence is mitigated by the growth technique (see also  Ref. \cite{Rajac23}).
\\
To provide greater clarity to the observed effect, it is useful to try to estimate the components of the magnetization in Si.
As stated above, the theoretical XMCD transition from a L$_{2,3}$ edge involves the core level 2p of the chemical species.
In the present case of silicon, further simplifications can be employed; the transition can in fact be considered probing only the empty s-band, as the contribution from the d-band to the conduction band arises solely at higher energies.
Accordingly, in the range 0-3 eV above the absorption edge any contribution other than from a s-band can be neglected \cite{Bianconi87}.
The theoretical XMCD signal is in general comprised of three terms: the orbital angular momentum $\langle L_z \rangle$, the spin angular momentum $\langle S_z \rangle$ and the magnetic dipole momentum $\langle T_z \rangle$.
However, as the L$_{2,3}$ transition arises from a p- to a s-band, further simplifications may be applied.
The final state in fact doesn't allow a nonzero angular momentum, as well as the magnetic dipole momentum \cite{Wu94}, if any contribution arising from surfaces and interfaces is excluded.
As a result, the spin sum rule from the initial filled band $c$ to the final band $l$ with occupation $n$ can be written as \cite{Altarelli98, Carra93}

\begin{multline}
\frac{\int_{j^+} dE (I^+ - I^-) \ - \ \frac{c+1}{c} \int_{j^-} dE (I^+ - I^-)}{\int_{j^ + + j^-} (I^+ + I^0 + I^- )} = \\
 \frac{l(l+1) \ - \ c(c+1) \ - \ 2}{3ch} \langle S_z \rangle
\end{multline}

Where the labels $c$ and $l$ refer to the initial filled core level and the final $h$-hole filled level respectively and the energy ranges $j^+$ and $j^-$ pertain to the transition to a state with final moment $c+1/2$, or L$_3$, and $c-1/2$ or L$_2$, respectively.
Since the initial and the final states are a p-band (c = 1) and a s-band (l = 0), whereas the holes $h$ can be determined as $4l + 2 - n$ ($n$ being the number of electrons in the band), the expression can be further reduced to

\begin{multline}
\frac{\Delta A_3 \ - \ 2 \cdot \Delta A_2}{3 A} = - \frac{2}{3} \langle S_z \rangle \ \ \ \ \ \ \ \ \ \ \ \ \ \ \ \ \ \ \ \ \ 
\end{multline}

Where $\Delta A_3$ and $\Delta A_2$ are the XMCD asymmetries integrated on the L$_3$ and L$_2$ edges respectively and $A$ the reflectivity edge integrated across the whole L$_{2,3}$ edge.
\\
To apply the spin selection rule, it is necessary to unambiguously separate the two absorption edges.
However, according to the literature, this is not possible when the edges overlap, as occurs at the M$_{2,3}$ edge for magnetic transition metals, as well as at the Si L$_{2,3}$ edge.
Despite this challenge, the XMCD signal in reflection at the Si L$_{2,3}$ edge displays a clear step where the contribution from the L$_2$ edge begins to increase.
Based on the assumptions outlined in the previous paragraph, the L$_3$ and L$_2$ contributions to the total XMCD signal were estimated.
The XMCD signal was fitted with two step-functions separated by the Si spin-orbit splitting of 0.61 eV and each aligned at higher energies with the same exponential decay, which accounts for the reduced magnetic sensitivity as one moves away from the Si edge.
Subsequently, the amplitudes of the two components were normalized to the mean reflectivity signal in the range up to 3 eV above the edge.
The normalized values for the L$_3$ and L$_2$ XMCD components can be found in Table~\ref{tab2}.
From these values it is possible to extract the spin angular momentum $\langle S_z \rangle$ for LD and HD, which are respectively $-0.0295 \pm 0.0002 \ \mu_B$ and $+0.0174 \pm 0.0005 \ \mu_B$.
These values are comparable to the spin magnetic moment measured at Si L-edge on Si-based Heusler alloys \cite{Antoniak12, Emmel14}.
\\

\begin{table}[ht!]
\captionsetup{justification=centerlast}
\begin {center}
\begin{tabular}{c D{,}{.}{-1} D{,}{.}{-1}}\toprule
& \multicolumn{1}{c}{\ \ \ \textit{Si L$_3$}} & \multicolumn{1}{c}{\ \ \ \textit{Si L$_2$}} \\ \midrule
LD  & -0 , 0230 & -0 , 0410 \\
HD & +0 , 0367 & +0 , 0357 \\ \bottomrule
\end{tabular}
\caption{\label{tab2} 
XMCD L$_2$ and L$_3$ components of the fits for LD and HD normalized per the mean reflectivity edge $A$ in the range 0-3 eV above the Si L$_{2,3}$ edge.
}
\end {center}
\end{table}

The spin and orbital angular momentum in Ni can be obtained by theoretical and experimental literature data taken at the Ni L-edges.
The values of the measured magnetic moments deviate considerably from the bulk values only for films of 2 ML or less \cite{Dhesi99}.
As in the present case the thickness of the Ni film is far greater than 2 ML, it is possible to fairly estimate the magnetic moments as the bulk values ones \cite{Schaefer92}, that is 
$\langle S_{z (Ni)} \rangle$ = 0.47 $\mu_B$ and $\langle L_{z (Ni)} \rangle$ = 0.05 $\mu_B$ (the case of a nanocrystalline Ni sample has been considered, which seems a fair assumption based on the HRTEM data).
\\
Our finding illustrate the feasibility of observing extremely weak induced magnetic moments in heterostructures. 
Further, by evaluating the induced magnetic moment it can also provide acceptance to the abovementioned mechanisms.

\section{Conclusions}

In this work, the sistematic investigations of the XMCD signals at the Ni M$_{2,3}$ and Si L$_{2,3}$ edges in two distinct Ni/Si interfaces were carried out. The samples, differentiated solely by the doping level of the substrate, demonstrate a nonzero magnetization of the silicon substrate due to the magnetic proximity effect from the Ni film. Remarkably, the sign of the magnetization in the two cases is found to be opposite; the low-doped sample aligns parallel to the magnetic film, while the high-doped sample aligns antiparallel.

This observed discrepancy in magnetization states suggests the presence of an intricate competition among various mechanisms responsible for the magnetization reversal. The two plausible mechanisms that could give rise to the magnetization are put forward: either through the tunnelling of spin-polarized electrons from the Ni magnetic layer or due to the modification of the exchange splitting of electrons in the silicon depletion layer, induced by the effective magnetic field.

To fully elucidate the behaviour of the substrate in these systems and the role of the magnetic proximity effect, further theoretical investigations are warranted. The insights gained from this study contribute to the comprehension of the complex interplay between ferromagnetic materials and semiconductors at the interface. This knowledge paves the way for the development of innovative spintronic devices that harness the unique properties of these hybrid systems.

\bibliographystyle{apsrev4-2}
\bibliography{bibliography_XMCD}

\end{document}